# The Substrates of Integrated Neurocognitive Rehabilitation Platforms (INCRPs)

Iman Ghodratitoostani[1,2], Ali-Mohammad Kamali[3,4,5], Mahshid Tahamtan[3,4,5], Neda Mohammadi[3,4,5], Hadi Aligholi [3,4,5], Mohammad Nami[3,4,5,6*]

[1]Neurocognitive Engineering Laboratory (NEL), Center of Engineering Applied to Health, University of São Paulo, São Carlos, Brazil
[2]Reconfigurable Computing Laboratory, Institute of Mathematics and Computer Sciences, University of São Paulo, São Carlos, Brazil
[3]Neuroscience Laboratory (Brain, Cognition and Behavior), Department of Neuroscience, School of Advanced Medical Sciences and Technologies, Shiraz University of Medical Sciences, Shiraz, Iran
[4]Department of Neuroscience, School of Advanced Medical Sciences and Technologies, Shiraz University of Medical Sciences, Shiraz, Iran
[5]DANA Brain Health Institute, Iranian Neuroscience Society, Fars Chapter, Shiraz, Iran
[6]Academy of Health, Senses Cultural, Sacramento, CA, USA

*Corresponding author:*
Mohammad Nami MD, PhD, Department of Neuroscience, School of Advanced Medical Sciences and Technologies, Shiraz University of Medical Sciences, Shiraz, Iran. torabinami@sums.ac.ir

## Abstract

The integrated neurocognitive rehabilitation platforms (INCRPs) refer to infrastructures and teams integrated for set of interventions which aim to restore, or compensate for cognitive deficits. Cognitive skills may be lost or altered due to brain damage resulting from diseases or injury. The INCRP is a two-way interactive process whereby people with neurological impairments work with specialists, professional staff, families, and community members to alleviate the impact of cognitive deficits. This perspective paper would highlight key elements required in INCRPs.

## Background

Areas of cognitive deficits may include attention, memory, mental flexibility, learning, language, perception, sensory-motor integration and other executive functions (e.g., planning, organization) (1). Although some endeavors have been in place to establish integrated neurocognitive rehabilitation platforms (INCRPs), there still remain gaps to be bridged (2).

The INCRP largely depend on well-integrated knowledge of the brain networks and architecture ranging from neurofunctional-cognitive models, multimodality head models, virtual reality and clinical information systems to clinical decision support systems (3, 4).

INCRPs are currently lacking in many research and clinical setups due to the fact that applied neuroscience is just in the prime of its journey for being positioned as a key element in healthcare infrastructures to offer system designs and work-flow benchmarks for future brain-health services aiming to improve the quality and sustainability of brain and cognitive productivity at societal level (5, 6).

Our shared initiative is establishing three INCRPs including the neuroscience laboratory (NSL) Neurocognitive Engineering Laboratory (NEL) and the Brain Health Institute ([www.danabrain.ir](www.danabrain.ir)) has inspired to extend the working-team concept and collaborative network in investigating and remadiating a range of neurodevelopmental, cognitive, behavioral and neurofunctional predicaments (7). The INCRP developed at our setting comprises platform for the assessment and remediation of cognitive impairments and neurobehavioral dysfunctions in a range of neuropsychiatric disorders(8). The sleep health and sleep disorders unit has also been a platform to comprehensively assess and improve sleep related issues in relation with neurocognitive complaints (9). In addition, the BrainCheck initiative has pursued a full range of brain check-up evaluations over a 4-hr process as an evidence-based screener for brain health and cognitive fitness (10).

The BrainCheck initiative at our setup has strived and continues to put together efforts to establish an INCRP based on key substrates including a proper platform for EEG signal processing and source localization, image processing, segmentation and functional brain mapping, applications for co-registration of anatomical and functional brain images, Bio-physical model of current distribution in the brain, neurofunctional-cognitive modeling and brain network connectivity, multimodality head model and to formulate high performance computational platform (Neural Coding, Neural Circuit Dynamics, and Neuromodulation) (11, 12).Over and above, developing transcranial electrical stimulation (tES) protocols in cognitive rehabilitation as well as virtual reality and 3D games in cognitive remediation have possibly been critical to such efforts (13-15).

This would, in long-term, inspire to progress in methodology and level/quality of diagnostic and interventional services beyond the issue of state-of-the-art technological structure (16, 17). That is the creativity and integrity of the multidisciplinary team which drives this process towards new frontiers (18). Below is a proposed overview on what need to be pursued in INCRPs.

### *Proper platform for EEG signal processing and source localization*

In addition to the ordinary signal processing methods used in research and clinical practice setups, there are some sophisticated methods which need to be developed and used in signal processing (19). Moreover, there are some results which need to be interpreted by signal processing and biostatistics experts (20).

The electroencephalographic (EEG) signal processing are expected to provide key insights for the evaluation, monitoring and rehabilitation of cognitive disabilities (21, 22). EEG signals should be treated in a professional manner so that necessary algorithms are designed for the evaluation of a cognitive disability (23). In addition, they should be comprehensively analyzed not only to facilitate the development of neurofunctional models, but also to validate them (24).

The tasks which correspond to such an objective are selected based on the clinicians', experimentalists' and neurocognitive modelers' needs (25). Such tasks are categorized as: 1-development of algorithms based on EEG upon evaluating cognitive responses, 2-denoising biomedical signals in the presence of electrical stimulation, network processing (e.g., producing maps, causality processes, etc.) to characterize the propagation of information in a cortical region of interest (26-28).

### *Applications for co-registration of anatomical and functional brain images*

Co-registration of anatomical and functional brain images will help defining an imaging application based on the photogrammetry system for EEG sensors localization for different types of existing EEG caps in the market (29). This technology would be used to identify the 3D position of each EEG sensor on the scalp (30). The sensor coordinate map is important for increasing the accuracy of electrical source imaging, relative to the default positions, as it describes the true position of the EEG sensors relative to the whole head model (31). This capability improves the accuracy of source localization performed in the head model software (32).

This would plan to develop photogrammetry applications to capture images of all the sensors on the head in a panoramic style (33). This is expected to provide convenience for researchers, clinicians, and their research participants and patients (34). The photographs provide a permanent record of the sensor positions for a given exam, allowing for verification and analysis at any time after the exam is completed (35).

The improving image processing methods on online and desktop applications would be expected to enable automatic identification of sensors from the images (36). As such, users can quickly and more accurately derive 3D sensor positions which adds advanced 3D head model visualization (37). This is known to permit conforming head models to individual head geometry, and the capability for using individual MR images to create an individual head model (38, 39).

### *Bio-physical model of current distribution in the brain*

When an electrical stimulation paradigm is presented, it is desired to know what would be the current/ electric field distribution inside the brain especially at cognitively related networks. Bio-physical model of current distribution in the brain can tells us to what extent a given brain area has been affected by the externally applied electrical current (40, 41).

Having this information, one can determine for instance which multi-site configuration is necessary for the multi electrode stimulation.

Shapes of stimulating electrodes, stimulation phases, frequencies and amplitudes can be better controlled by having a biophysical model of current distribution resulting from a single or, multi-site, or an array laminar electrodes (42).

For this objective, the MRI of a given subject or patient should be processed so that a high-resolution mesh is created (43, 44). The propagation of the current and/or electric field is found by solving Maxwell equations for the high-resolution mesh (45, 46) . Current distribution for the regions of interest should be visualized in a proper manner.

### *Neurofunctional-cognitive modeling and brain network connectivity*

One of the requirements to design a novel and efficient cognitive rehabilitation method is to have a proper definition for the cause(s) of a given cognitive predicament. One of the advantages of the multi-disciplinary approach is that the theoretical modeling aspects of cognitive disorder are also considered.

In favor of developing better new realistic neurofunctional-cognitive models, our understanding about underlying mechanisms of cognitive disorder will be increased (47). Later, clinical scientists may develop models according to which cognitive aspects of the disorder are better described.

Progress in neurofunctional-cognitive modeling will shed lights on cognitive disorder rehabilitation since such models will be able to describe which brain area is more engaged, which types of plasticity have been created, and which reinforcements may increase or decrease the severity of the cognitive predicament.

In order to validate a neurofunctional-cognitive model, various types of clinical and experimental tests are needed. In terms of clinical tests, psychological cognitive tests should be performed in clinics. In addition biomedical signals originating from the brain activity should be recorded and processed in cognitive and psychiatric related patients. In addition, functional Magnetic Resonance Imaging (fMRI) can provide supplementary information in this regard.

### *Multimodality head model and executive high performance computational platform (Neural Coding, Neural Circuit Dynamics, and Neuromodulation)*

Neural coding and neural circuit dynamics are conceptual foundations upon which mechanistic understanding of the brain is pursued (48-50). At the microscopic scale, the brain consists of vast networks of neurons that are wired together with synaptic connections to form neural circuits (51, 52). In an active brain, each neuron can have electrical and chemical activity different from that of the neighboring circuit; thus some neurons can play specialized roles in different tasks (53). Yet, the activity of each neuron also depends on that of the others in the circuit through the synaptic connections which define the circuit's architecture (54). Synaptic connections can change strength as a result of recent activity in the circuit, meaning that circuit architecture is constantly modified by experience (55). A thinking brain can therefore be viewed as an immensely complex pattern of activity distributed across multiple, ever-changing circuits.

Neural coding refers to how information about the environment, the individual's needs, motivational states, and previous experience are represented in the electrical and chemical activity within a neural circuit (56). Elucidating the nature of complex neural codes and the logic underlying them is among central goals in INCRPs.

As different neurons become silent or active in a thinking brain, the pattern of activity shifts in space and time across different circuits in various brain regions. These shifting patterns are referred to as neural circuit dynamics. A key to understanding how the brain works is to determine how the neural dynamics across these vast networks process information in relation to behavior (57). For instance, what is the form of neural dynamics in a circuit which underlay a decision? What are the dynamically changing patterns of activity when processing a sentence or predicting a future action?

Considering the fact that the basic electrophysiological properties of single neurons are common across brain areas and species, it is likely that some, if not many, fundamental forms of neural dynamics would generalize across investigational models (58) . One goal of the cognitive rehabilitation platform is identify and characterize the universal forms of neural circuit dynamics, likely represented by dynamical motifs such as attractors, sequence generation, oscillation, persistent activity, synchrony-based computational analyses.

The above analyses need to be balanced with rapid flow of information that drives cognition, perception. Meanwhile, actions are slower modulatory influences associated with arousal, emotion, motivation, physiological needs, and the circadian states. In some cases, these slower influences are associated with specialized neuromodulatory chemicals i.e. neuropeptides, often produced deep in the brain or even in peripheral tissues, that can act locally or globally to change the flow of information across other brain circuits (59) .

Indeed, the neuromodulatory modifications of synaptic strength can 'rewire' a circuit to produce different dynamic patterns of activity at different points in time. As such, the cognitive rehabilitation platform should strive for a deeper understanding of these powerful but elusive regulators of mood and behavior.

### *Developing transcranial electrical stimulation (tES) protocols in cognitive rehabilitation*

A malfunctioning brain in terms of dysregulated data processing might result in cognitive predicaments. In that perspective, two treatment approaches would be regulating bottom-up and top-down processes(60, 61). For bottom-up rehabilitation, electrical, acoustic and visual stimulation are performed(62-64). For the top-down approach, effects of transcranial or deep current stimulation of the cognitive network have been investigated in several studies(65-67). Transcranial electrical stimulation may be comprise direct- (tDCS) or alternative current stimulation (tACS)(68, 69).

Transcranial- and deep electrical stimulations can reduce or increase the activity of high-level-processing brain areas (70-72). Deep electrical stimulations will be performed by needle electrodes (single or laminar) (72). Transcranial stimulations are typically performed by oddball electrodes (73).

The obtained dataset (after being stored in a popularized databank format) will be analyzed in different aspects. Safety issues, model validation based on the current source density maps, propagation of information in cognitive-disordered and in healthy brains are the topics envisaged (74). The results obtained would also be beneficial in guiding clinical experiments as well.

### *Developing Virtual reality and 3D games in cognitive rehabilitation*

Cognitive and psychiatric disorders is caused by intrinsic or extrinsic factors which may result in various disabilities such as motor, sensory, behavioral, or cognitive dysfunction depending on the brain area affected (75). Cognitive impairment due to disorders is an important factor affecting patients' independent functions and participation in activities of daily living (76). It can also influence motivation and the ability to participate in rehabilitation programs (76).

Therefore, for successful rehabilitation, accurate and comprehensive cognitive assessment and treatment formulation are required. For cognitive rehabilitation of patients, traditional treatment and computer-based cognitive therapy are primarily used (77). Virtual reality (VR) technology is gaining recognition as a useful tool for cognitive research, evaluation, and rehabilitation (78). VR systems allow users to interact in various sensory environments and to obtain real-time feedback on their performance using computer technology. The virtual environment offered via VR technology makes it possible for patients to participate in activities in settings and environments similar to those encountered in real life(79).

In addition, the VR tools can be used to record accurate measurements of the subject's performance (80). VR has been used as a tool to diagnose cognitive impairments and as an auxiliary tool to formulate new treatments. Although the use of VR in cognitive rehabilitation has been increasing, few systematic reviews have investigated the use of VR programs in cognitive rehabilitation (81). Studies using VR programs for cognitive intervention were reviewed according to PICO (patient, intervention, comparison, and outcome) method as the main approach in evidence-based rehabilitation protocol development (82).

Today, we are experiencing a change in our attitudes regarding deficiency and disability. As the population grows older, the need of new therapies to treat the cognitive decline which occurs with age becomes more urgent. The popularization of computers has opened the door to solutions for cognitive rehabilitation based in Information Technology (IT). Games that were previously designed exclusively for entertainment are now considered as valuable tools for "brain-training" (83). Currently, two complementary cognitive interventions take advantage of them. "Extrinsic" cognitive training uses computer games to exercise certain skills that might influence other mental functions (84). In contrast, "intrinsic" interventions utilize the game as part of a neurofeedback protocol, to visualize and promote plastic changes in the brain (85).

With the concurrent multivariate analysis of EEG signals based on graph theory (86), we might be able to describe the current brain state in real-time while the subject is receiving a cognitive intervention using a computer game. This information is used to calculate a Cognitive Ability Score (CAS), which serves as reference value for the training (87). The system can be used both at the hospital and at home to allow regular training and provide constant communication with the therapists. Tools such VR can be extremely useful in the near future to prevent or remediate cognitive predicaments (82).

### *Developing decision support system cognitive rehabilitation planning*

Clinical management system (CMS) entered the medical systems during the last decades with significant advances, including noticeable improvements in medical care, starting from ease of data storage and access of digital imaging through computerized medical data, accessing on-line literature, patient monitoring, therapy planning and a support system for medical diagnoses.

The CMS together with content-management systems are shown to be valuable tools to help practitioners when facing challenging medical problems i.e. differential diagnoses.

A main objective within the INCRPs is to develop novel general cognitive disorder-clinical management system (CDCMS).

Uncertainty in decision making with regards to treatment options is an accepted fact of life for clinicians and automated systems. The INCRPs are aimed to manage or reduce uncertainty, or to make it explicit. In reality, diagnostic approaches and treatments are subject to revisions and possible adjustments.

### *Conclusive remark*

The key element to help INCRPs flourish is to progress beyond the-state-of-the-art technology and materialize the creativity of a cross-disciplinary team to develop and implement novel, yet evidence-based, approaches in diagnosis and remediation of neurocognitive predicaments. Moving from generic approaches in neurocognitive intervention to predictive medicine and precision medicine would be expected to optimize neurocognitive health and to improve treatment outcomes. This would be also expected to leave an impact on the cost-utility balance of services offered in the field of neuropsychiatry, clinical neuroscience and cognitive medicine.

### *Disclosure*

None